\documentclass[12pt]{article} \usepackage{amsmath} \usepackage{amssymb}
\usepackage{epsfig}
\usepackage{textcomp}

\newcommand{\cA}{\mathcal{A}}
\newcommand{\cB}{\mathcal{B}}

\newcommand{\cP}{\mathcal{P}}
\newcommand{\cR}{\mathcal{R}}
\newcommand{\cX}{\mathcal{X}}
\newcommand{\cY}{\mathcal{Y}}

\newcommand{\cO}{\mathcal{O}}

\newcommand{\oS}{{\hbox{\openface S}}}

\font\openface=msbm10 at10pt

\begin{document}
\title {Confounding Causality Principles: comment on R\'edei and san Pedro's ``Distinguishing Causality Principles''} \author{Joe Henson\footnote{email: j.henson@imperial.ac.uk} } \maketitle
\begin{abstract}
R\'edei and san Pedro discuss my ``Comparing Causality Principles,'' their main aim being to distinguish reasonable weakened versions of two causality principles presented there, ``SO1'' and ``SO2''.  They also argue that the proof that SO1 implies SO2 contains a flaw.  Here, a reply is made to a number of points raised in their paper.  It is argued that the ``intuition'' that SO1 should be stronger than SO2 is implicitly based on a false premise.  It is pointed out that a similar weakening of SO2 was already considered in the original paper.  The technical definition of the new conditions is shown to be defective.  The argument against the stronger versions of SO1 and SO2 given by R\'edei and san Pedro is criticised.  The flaw in the original proof is shown to be an easily corrected mistake in the wording.  Finally, it is argued that some cited results on causal conditions in AQFT have little relevance to these issues, and are, in any case, highly problematic in themselves.
\end{abstract}
\vskip 1cm

In \cite{Henson:2005wb} I presented a framework within which some problems surrounding Bell's local causality condition and Reichenbach's Principle of Common Cause (PCC) could be addressed.  The main aim was to clarify the relation between variants of the causal condition that is used in the derivation of the Bell inequalities.  The central idea behind the more technical side of the work was the following: that various conditions in the literature claimed as alternatives to Bell's are either equivalent, or stronger in the sense that they incorporate inessential extra restrictions, or could be shown to be inadequate in a clear way.  In this connection, the main results in the paper strengthen the justification of Bell's causal condition, and defend the significance of Bell's theorem against the possibility of replacing local causality with some weaker alternative.

Central to the discussion are two causality principles that can be defined in that framework, SO1 and SO2.  Essentially, the two conditions are as follows (\cite{Henson:2005wb} contains more exact definitions, and the definitions of other terms used below).  Consider two spacelike separated regions $\cA$ and $\cB$ in a spacetime $\oS$.  A ``screening off'' type condition states that any pair of probabilistic events $A$ and $B$ occurring in any such regions $\cA$ and $\cB$ respectively will be uncorrelated once probabilities are conditioned on any ``full specification'' of events in the relativistic past.  The past region can be defined in different ways.  For SO1 the region is the ``mutual past'' $\cP_1=J^-(\cA) \cap J^-(\cB)$, where $J^-(\cA)$ is the causal past of $\cA$, while SO2 uses the larger ``joint past'' region $\cP_2=J^-(\cA) \cup J^-(\cB) \backslash (\cA \cup \cB)$, where $\backslash$ is set difference.  In \cite{Henson:2005wb}, proofs are provided that these conditions are equivalent.

R\'edei and san Pedro \cite{Redei:2012} take issue with the equivalence result using two main arguments.  Firstly, they state that the ``equivalence seems counterintuitive'' in that SO2 should intuitively be weaker than SO1, and cite results on causal principles in AQFT that seem consistent with the intuition.  The proposed solution is that SO2 should be replaced by a similar but weaker condition.  Secondly they find fault with the proof that SO1 implies SO2.

\paragraph{Simpson's paradox repeated.} The common intuition that SO2 must be the weaker condition is based, it seems, on the idea that conditioning on \textit{more} events in the past can only make \textit{more} pairs of events uncorrelated, not less.  If this was true then SO1 would easily imply SO2, but equivalence would be in doubt.  This reasoning is incorrect, however.  Uncorrelated events can become correlated once probabilities are conditioned on past events, leading to the so-called ``Simpson paradox'' \cite{Uffink:1999}.  Because of this, if arguments that SO2 implies SO1 seem endangered by causes of future correlations lying outside $\cP_1$, then arguments that SO2 implies SO1 should seem equally endangered by a Simpson's ``paradox'' brought on by events outside $\cP_1$.  An argument along these lines is touched on in \cite{Henson:2005wb} before the equivalence proof is given.

\paragraph{Finite SO2 and SO2w.} The weakened conditions proposed by R\'edei and san Pedro are called ``finite SO1'' and ``finite SO2''.  The weakening consists in restricting the set of regions to which the condition applies, allowing correlations between events in certain spacelike regions that are not subject to the screening off condition.  The regions to which finite SO1 and finite SO2 do not apply are ``causally infinite'' regions $\cA$ such that the causal closure of $\cA$ contains its own causal past (see \cite{Redei:2012} for details).  This is claimed to rule out regions like $J^-(\cB) \backslash J^-(\cA)$ which would, if correct, prevent the equivalence proofs in \cite{Henson:2005wb} from applying to the new finite conditions.

A similar variant of SO2, which is not mentioned in \cite{Redei:2012}, was considered in section 2.4 of \cite{Henson:2005wb}.  There SO2w is defined, in which the regions $\cA$ and $\cB$ are restricted to be regions that ``are of finite extent and do not contain any part of the initial hypersurface''.  R\'edei and san Pedro's finite SO2 condition has the advantage that it is defined purely with respect to the causal relation on the the set of spacetime points, which in my original paper was the only structure explicitly given to the ``spacetime'' $\oS$, whereas mention of ``finite extent'' implicitly uses more structure.  Finite SO1 is also new.  Beyond that, the conditions are clearly very similar, at least in intent.

\begin{figure}[ht]
\centering \resizebox{3in}{2.5in}{\includegraphics{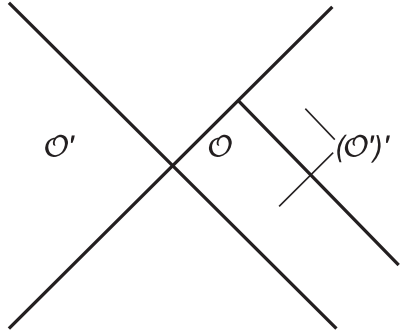}}
\caption{\small{
A spacetime diagram showing the problem with the definition of causal finiteness.  The region $\cO$ corresponds to the region $\cB \cup \cY$ in Figure 1 of \cite{Redei:2012}, which the authors claim to be causally finite.  The diagram shows the causal complement of this region $O'$, and the causal closure $(O')'$.  It is easy to see that the causal closure $(O')'$, which is the quadrant to the right in the diagram, does not contain its own past.
\label{f:finiteness}}}
\end{figure}

\paragraph{``Causal finiteness'' is defective.}  However, the definition of the term ``causally finite'' does not do the intended job, even in the intuitive case referred to in \cite{Redei:2012}, that is, 2D Minkowski space.  The problem is illustrated in figure \ref{f:finiteness}.  Let us use light-cone co-ordinates $u=t+x$ and $v=t-x$ in this space.  Consider a region defined by $u \geq 0$, $v \leq 0$ and $u \leq u*$ where $u*$ is positive.  This region corresponds to the region $\cB \cup \cY$ in Figure 1 of \cite{Redei:2012}, and for brevity it will be called $\cO$ here.  It extends infinitely into the past in an obvious sense.  R\'edei and san Pedro claim that such regions are also causally infinite in the sense of their definition.  However, the causal complement $O'$ is the set of all points such that $u < 0$ and $v > 0$, and so the causal closure $(O')'$ is the quadrant defined by $u \geq 0$, $v \leq 0$.  This causal closure region $(O')'$ does \textit{not} contain its own past, and so $O$ is causally \textit{finite} by their definition.  Thus, in 2D Minkowski space at least, the proof of the equivalence of SO1 and SO2 actually does carry over to the ``finite'' conditions!  This undermines the motivation for the definition of ``Causal finiteness''.

It is certainly worthwhile to consider the consequences of modifying SO1 and SO2.  But before analysing ``finite'' SO1 and SO2, we first need to improve on the definitions.  When this task is completed, the more substantial questions can be dealt with.

\paragraph{Motivation for finite SO1 and SO2 is problematic.}  However, even given that this technical problem can be resolved, the motivation given for preferring the alternative conditions is questionable.  Briefly, R\'edei and san Pedro state that a causally infinite region ``might very well be empirically inaccessible in its entirety... hence one might not be in the position to decide empirically whether [an event in that region] has happened or not''.  Hence events in such regions should not be subject to screening-off conditions, they argue.

This argument depends on a strong epistemic interpretation of the idea of ``domains of decidability'' from \cite{Henson:2005wb}, that is, that they should be understood as the regions to which we would have to have access in order to know whether the event had occurred or not.  Butterfield (\cite{Butterfield:2007a}, footnote 3, p.6) makes a compelling criticism of the use of epistemic language in \cite{Henson:2005wb}.  It is, he states, a ``gloss'' that would be unnecessary, were it not an attempt to avoid the need to bring in the difficult notion of events being ``intrinsic'' to regions (an attempt which in any case fails).  If we agree, and accept that ``domains of decidability'' should not be thought of epistemically, the argument given above cannot be carried through.  There is a further problem with the argument: R\'edei and san Pedro's finite SO1 and SO2 conditions still allow events whose least domain of decidability is causally infinite to be conditioned on, but they have argued against causally infinite least domains of decidability in general.

\paragraph{SO1 does imply SO2.}  In section 5 of their paper R\'edei and san Pedro challenge the proof that SO1 implies SO2.  Rather than a being substantial error, however, this boils down to an erroneous restatement, in the explanation of the proof, of a result proved earlier in the paper.  Correcting the restatement completes the proof.

The first part of the proof in \cite{Henson:2005wb} shows that, if we assume SO1, any pair of events $A$ and $B$ in any spacelike regions $\cA$ and $\cB$ will be uncorrelated when probabilities are conditioned on any full specifications of three particular disjoint regions, the union of which is the joint past $\cP_2$. The last part of the proof offered is the following: ``[f]rom corollary 1, if $C \in \Phi(\cP_1)$, $X \in \Phi(\cX)$ and $Y \in \Phi(\cY)$ then $C \cap X \cap Y \in \Phi(\cP_2)$'' where $\Phi(\cA)$ refers to the set of full specifications of $\cA$.  That is, if we condition on any full specification of the three regions, that is equivalent to conditioning on some full specification of their union, and this is claimed to complete the proof.  R\'edei and san Pedro rightly point out that this is not the same as saying that \textit{any} full specification of $\cP_2$ is equal to the union of some such triple of full specifications $C$, $X$ and $Y$ for the corresponding disjoint regions, which is what we need to show that SO2 holds.  However, when we return to corollary 1 to see what is actually says, we discover that it does indeed complete the proof.  Corollary 1 is as follows:  ``if $\cR=\bigsqcup_i \cA_i$ for some finite set of regions $\cA_i$, then a full specification $F$ of $\cR$ can be written $F=\bigcap_i A_i$ where $A_i$ is a full specification of $\cA_i$''.  Here $\sqcup$ signifies a disjoint union.  The meaning is that \textit{any} full specification $F$ of $\cR$ can be written in this way.  This corollary is saying exactly what R\'edei and san Pedro point out is missing from the proof.  The error, such as it is, consists in explaining the content of corollary 1 incorrectly.  So while the objection helps to clarify the proof, the problem is not substantial.

Thus the equivalence between the original forms of SO1 and SO2 is not actually in danger.  However, the conjecture that finite SO1 implies finite SO2 is also referred to as ``likely true and intuitively plausible'' in section 5 of \cite{Redei:2012}, while the opposite implication is considered less likely to hold. An intuition refined in the face of Simpson's paradox should find no obvious reason to consider one of the implications more likely to be true than the other (even ignoring the problems with the definition of casual finiteness).

\paragraph{Results from AQFT have limited relevance.} R\'edei and san Pedro go on to review some results on causal conditions in algebraic quantum field theory (AQFT), claiming that they provide some evidence for relations between the screening off conditions, for example that finite SO2 is strictly weaker than the original SO2.  It is misguided to compare results based on AQFT to results in the framework presented in \cite{Henson:2005wb} in this way, for a number of reasons.

Not only the comparison, but also the general significance of the cited results, is questionable.  Citing \cite{Redei:2001yn} in section 5 of their paper, R\'edei and san Pedro state that if AQFT satisfies a certain condition, ``the \textit{finite} SO2 holds in AQFT'' in ``a specific sense'' which they then specify.  The causal principle referred to is a statement in the AQFT framework, analogous in form to Reichenbach's original statement of the Principle of Common Cause (definition 2 in \cite{Redei:2001yn}), with the relativistic joint past as the definition of the past region.  It is surprising to see this referred to as a version of SO2, in any sense, in a paper based on a detailed reading of \cite{Henson:2005wb}.  The first half of \cite{Henson:2005wb}, leading up to the definitions of SO1 and SO2, deals explicitly and at length with this condition. The original PCC condition is contrasted in detail with screening off conditions.  The conclusion is reiterated here to avoid any ambiguity: Reichenbach's original PCC suffers from numerous inadequacies when applied in this manner.  Doing so leads to absurdities: this principle rules out many manifestly causal situations and is vulnerable to Simpson's ``paradox''.  What R\'edei and Summers have shown is that something formally analogous to this inadequate condition holds in AQFT.  This should only cause us to question whether the formal analogy, or anything like it, captures anything physically interesting.  It does not tell us anything about SO2, finite or infinite.

Beyond this particular work, there are broader reasons for caution in comparing results in the AQFT framework to that in \cite{Henson:2005wb}.  The frameworks are far from equivalent.  Furthermore, definitions of causal principles analogous to SO1 and SO2 are not settled in the AQFT framework, and different choices can lead to different results as to whether AQFT obeys various causal principles (see for example \cite{Butterfield:2007a}).

The underlying problem is that, in the SO1 and SO2 conditions, and other such conditions such as Stochastic Einstein locality \cite{Butterfield:2007a} and Bell's own formulations, the ``events'' being conditioned on can be said to describe something like Bell's ``beables'' \cite{Bell:1987}.  AQFT is, on the other hand, an axiomatisation of an operational theory.  In an important sense, any decision on what a full specification of past events should be in AQFT implicitly presupposes something about what the beables should be.  AQFT does not tell us what the beables should be.  It does not even tell us whether they should correspond to structures already in the AQFT framework or whether extra structures should be added.  For instance, in one case in which beables are added to a lattice QFT, the results are manifestly non-local \cite{Samols:1995zz}.  Thus any conclusions we can reach here about locality, meaning lack of superluminal influence, are inextricably tied to another problematic aspect of the story, namely what there is to influence or be influenced.

This should not be taken as an argument that defining conditions that are formally analogous to SO1 and SO2 in other frameworks is always useless (some speculation along these lines was made in \cite{Henson:2005wb} using the idea of the decoherence functional rather than the AQFT framework).  However, it is wrong to overinterpret such results as saying anything about the probabilistic framework of \cite{Henson:2005wb}, or for that matter, anything decisive about locality in the above sense.  To do the latter, it is at least necessary to make sure that the causal condition used is physically sensible.  This means taking on the arguments surrounding Bell's theorem more directly.

\bibliographystyle{plain}
\bibliography{refs}

\end{document}